\documentclass{pasj00}

\begin{document}
\SetRunningHead{Meng, Yang \& Geng}{WD+MS systems as the
progenitors of Type Ia supernovae with different metallicities}
\Received{2000/12/31}
\Accepted{2001/01/01}

\title{WD+MS Systems as the Progenitors of Type Ia Supernovae
   with Different Metallicities}

\author{X.-C. \textsc{Meng} %
  \thanks{We are grateful to the anonymous referee for his/her
constructive suggestions improving this manuscript greatly. This
work is supported by the Chinese National Science Foundation
(Grant Nos. 10603013).} and W.-M. \textsc{Yang}}

\affil{Department of Physics and Chemistry, Henan Polytechnic
University, Jiaozuo, 454000, China}\email{conson859@msn.com} \and
\author{X.-M. {\sc Geng}}
\affil{Academic publishing Center, Henan Polytechnic University,
Jiaozuo, 454000, China}

%

\KeyWords{stars: binaries: general---stars: supernovae: general---individual(SN 2002ic)---stars: white dwarfs} 

\maketitle

\begin{abstract}
The single-degenerate model for the progenitors of type Ia
supernovae (SNe Ia) is one of the two most popular models, in
which a carbon-oxygen white dwarf (CO WD) accretes hydrogen-rich
material from its companion, increases its mass to the
Chandrasekhar mass limit, and then explodes as a SN Ia.
Incorporating the prescription of Hachisu et al. (1999a) for the
accretion efficiency into Eggleton's stellar evolution code and
assuming that the prescription is valid for \emph{all}
metallicities, we carried out a detailed binary evolution study
with different metallicities. We show the initial and final
parameter space for SNe Ia in a ($\log P-M_{\rm 2}$) plane. The
positions of some famous recurrent novae in the ($\log P-M_{\rm
2}$) plane, as well as a supersoft X-ray source (SSS), RX
J0513.9-6951 are well explained by our model, and our model can
also explain the space velocity and mass of Tycho G, which is now
suggested to be the companion star of Tycho's supernova . Our
study indicates that the SSS, V Sge, is a potential progenitor of
supernovae like SN 2002ic if the delayed dynamical-instability
model in Han \& Podsiadlowski (2006) is appropriate.
\end{abstract}

\section{Introduction}
\label{sect:1}

Type Ia supernovae (SNe Ia) play an important role as cosmological
distance indicators and are successfully applied to determine
cosmological parameters (e.g. $\Omega$ and $\Lambda$;
\cite{RIE98}; \cite{PER99}), which have led to the discovery of
the accelerating expansion of the universe. However, the exact
nature of SNe Ia progenitors has not been clarified (see the
reviews by \cite{HN00}; \cite{LEI00}). It is generally agreed that
SNe Ia originate from the thermonuclear runaway of a carbon-oxygen
white dwarf (CO WD) in a binary system. The CO WD accretes
material from its companion, increases mass to its maximum stable
mass, and then explodes as a thermonuclear runaway. Almost half of
the WD mass is converted into radioactive nickel-56 in the
explosion (\cite{BRA04}), and the amount of nickel-56 determines
the maximum luminosity of SNe Ia (\cite{ARN82}).

According to the nature of the companions of the mass accreting
white dwarfs, two competing scenarios have been proposed, i.e the
double-degenerate channel (DD, \cite{IT84}; \cite{WI87}) and the
single degenerate channel (SD, \cite{WI73}; \cite{NTY84}). In the
DD channel, two CO WDs with a total mass larger than the
Chandrasekhar mass limit may coalesce, and then explode as a SN
Ia. Although the DD channel is theoretically less favored, e.g.
double WD mergers are more likely to result in accretion-induced
collapses rather than SNe Ia (\cite{HN00}), it is premature to
exclude the channel at present since there exists evidence that
the channel may contribute to a few SNe Ia (\cite{HOW06};
\cite{BRA06}; \cite{QUI07}). The single-degenerate Chandrasekhar
model is an alternative of DD and some surveys showed that many SD
systems are good candidates for SNe Ia (\cite{WI73}; \cite{NTY84};
\cite{PAR07}). In this model, the maximum stable mass of a CO WD
is $\sim 1.378 M_{\odot}$ (close to the Chandrasekhar mass,
\cite{NTY84}), and the companion is probably a main sequence star
or a slightly evolved star (WD+MS), or a red-giant star (WD+RG)
(\cite{YUN95}; \cite{LI97}; \cite{HAC99a, HAC99b}; \cite{NOM99};
\cite{LAN00}; \cite{HAN04}; \cite{MEN09a}). The SD model is
supported by many observations. For example, variable
circumstellar absorption lines were observed in the spectra of SN
Ia 2006X (\cite{PAT07}), which indicates the SD nature of its
precursor. \citet{PAT07} suggested that the progenitor of SN 2006X
is a WD + RG system based on the expansion velocity of the
circumstellar material, while \citet{HKN08} argued a WD + MS
nature for this SN Ia. Recently, \citet{VOSS08} suggested that SN
2007on is also possible from a WD + MS channel. In this paper, we
only focus on the WD + MS channel, which is a very important
channel for producing SNe Ia in our Galaxy (\cite{HAN04}).

Observationally, some WD + MS systems are possible progenitors of
SNe Ia (see the review of \cite{PAR07}). For example, supersoft
X-ray sources (SSSs) were suggested as good candidates for the
progenitors of SNe Ia (\cite{HK03b, HK03c}). Some of the SSSs are
WD + MS systems and some are WD + RG systems (\cite{DIK03}).
\citet{DIK03} reported that in every galaxy, there are at least
several hundred SSSs with a luminosity of $\geq10^{\rm 37}$ $\rm
erg$ $\rm s^{\rm -1}$ based on Chandra data from four external
galaxies (an elliptical  galaxy, NGC 4967; two face-on spiral
galaxies, M101 and M83; and an interacting galaxy, M51). They also
noticed that the SSSs appear to be associated with the spiral arms
in the spiral galaxies, which may indicate that SSSs are young
systems (WD + MS systems?). Recurrent novae may also be good
candidates as the progenitors of SNe Ia (\cite{BRANCH95}) and
several novae have been suggested to be possible progenitors
(\cite{HAC00a,HAC00b}; \cite{HK00,HK03a,HK05, HK06a, HK06b};
\cite{HKL07}). Some of the novae are WD +MS systems and some are
WD + RG systems.

A direct way to confirm the progenitor model is to search for the
companion stars of SNe Ia in their remnants. The discovery of the
potential companion of Tycho's supernova have verified the power
of the method and also the reliability of the WD + MS model (Tycho
G named in \cite{RUI04}). Recently, \citet{HERNANDEZ09} stressed
further the statement in \citet{RUI04} by analysing the chemical
abundances of Tycho G.

Many works have concentrated on the WD+MS channel. Some authors
(\cite{HAC99a, HAC99b, HKN08}; \cite{NOM99, NOM03}) have studied
the WD+MS channel by a simple analytical method to treat binary
interactions. Such analytic prescriptions may not describe some
mass-transfer phases, especially those occurring on a thermal
time-scale (\cite{LAN00},). \citet{LI97} studied this channel from
detailed binary evolution calculation, while they considered two
WD masses (1.0 and 1.2 $M_{\odot}$). \citet{LAN00} investigated
the channel for metallicities $Z=0.001$ and 0.02, but they only
studied the case A evolution (mass transfer during core hydrogen
burning phase). \citet{HAN04} carried out a detailed study of this
channel including case A and early case B (mass transfer occurs at
Hertzsprung gap, HG) for $Z=0.02$. Following the study of
\citet{HAN04}, \citet{MEN09a} studied the WD + MS channel
comprehensively and systematically at various Z and showed the
initial parameter spaces for the progenitors of SNe Ia and the
distributions of the initial parameters for the progenitors of SNe
Ia. Here, based on the study in \citet{MEN09a}, we want to show
the final parameter spaces of companions at the moment of SNe Ia
explosion and check whether the model used in \citet{MEN09a} can
explain the properties of some recurrent novae and SSSs, which are
suggested to be the possible progenitors of SNe Ia, and the
properties of Tycho G, which is the potential companion of Tycho's
supernova.

In section \ref{sect:2}, we simply describe the numerical code for
binary evolution calculations. The evolutionary results are shown
in section \ref{sect:3}. In section \ref{sect:4}, we briefly
discuss our results, and finally we summarize the main results in
section \ref{sect:5}.




\section{Binary Evolution Calculation}
\label{sect:2}

Our method for treating the binary evolution of WD + MS systems is
same to that in \citet{MEN09a}. In the following, we simply
redescribe our method. We use the stellar evolution code of
\citet{EGG71, EGG72, EGG73} to calculate the binary evolutions of
WD+MS systems. The code has been updated with the latest input
physics over the last three decades (\cite{HAN94}; \cite{POL95,
POL98}). Roche lobe overflow (RLOF) is treated within the code
described by \citet{HAN00}. We set the ratio of mixing length to
local pressure scale height, $\alpha=l/H_{\rm p}$, to 2.0, and set
the convective overshooting parameter, $\delta_{\rm OV}$, to 0.12
(\cite{POL97}; \cite{SCH97}), which roughly corresponds to an
overshooting length of $0.25 H_{\rm P}$. Ten metallicities are
adopted here (i.e. $Z=$0.0001, 0.0003, 0.001, 0.004, 0.01, 0.02,
0.03, 0.04, 0.05 and 0.06). The opacity tables for these
metallicties are compiled by \citet{CHE07} from \citet{IR96} and
\citet{AF94}. For each $Z$, the initial hydrogen mass fraction is
obtained from

 \begin{equation}
 X=0.76-3.0Z,
  \end{equation}
since this relation can well reproduce the color-magnitude diagram
(CMD) of some clusters(\cite{POL98}).

Instead of solving stellar structure equations of a WD, we adopt
the prescription of \citet{HAC99a} on WDs accreting hydrogen-rich
material from their companions. In a WD + MS system, the companion
fills its Roche lobe at MS or during HG, and transfers material
onto the WD. We assume that if the mass-transfer rate,
$|\dot{M}_{\rm 2}|$, exceeds a critical value, $\dot{M}_{\rm cr}$,
the accreted hydrogen-rich material steadily burns on the surface
of WD, and is converted into helium at the rate of $\dot{M}_{\rm
cr}$. The unprocessed matter is lost from the system as an
optically thick wind at a rate of $\dot{M}_{\rm
wind}=|\dot{M}_{\rm 2}|-\dot{M}_{\rm cr}$ (\cite{HAC96}). Based on
the opacity from \citet{IR96}, the optically thick wind is very
sensitive to Fe abundance, and it is possible for the wind not to
work when $Z< 0.002$ (\cite{KOB98}). Thus, there should be an
obvious low-metallicity threshold for SNe Ia in comparison with SN
II. However, this metallicity threshold was not found
(\cite{PRI07a}). Considering the uncertainties in the opacities,
we therefore assume rather arbitrarily that the optically thick
wind is valid for all metallicities.

The critical accretion rate is given by

 \begin{equation}
 \dot{M}_{\rm cr}=5.3\times 10^{\rm -7}\frac{(1.7-X)}{X}(M_{\rm
 WD}-0.4),
  \end{equation}
where $X$ is hydrogen mass fraction and $M_{\rm WD}$ is the mass
of the accreting WD (mass is in $M_{\odot}$ and mass-accretion
rate is in $M_{\odot}/{\rm yr}$, \cite{HAC99a}). The effect of
metallicities on equation (2) has not been included since the
effect is very small and can be neglected (\cite{MEN06}).

We adopted the following assumptions when $|\dot{M}_{\rm
2}|\leq\dot{M}_{\rm cr}$. (1) When $|\dot{M}_{\rm
2}|\geq\frac{1}{2}\dot{M}_{\rm cr}$, the hydrogen-shell burning is
steady and no mass is lost from the system. (2) When
$\frac{1}{2}\dot{M}_{\rm cr}>|\dot{M}_{\rm
2}|\geq\frac{1}{8}\dot{M}_{\rm cr}$, a very weak shell flash is
triggered but no mass is lost from the system. (3) When
$|\dot{M}_{\rm 2}|<\frac{1}{8}\dot{M}_{\rm cr}$ , the
hydrogen-shell flash is so strong that no material can be
accumulated by the accreting CO WD. Then, the growth rate of the
mass of the helium layer under the hydrogen-burning shell can be
defined as
 \begin{equation}
 \dot{M}_{\rm He}=\eta _{\rm H}|\dot{M}_{\rm 2}|,
  \end{equation}
where $\eta _{\rm H}$ is the mass accumulation efficiency for
hydrogen burning and its values is

 \begin{equation}
\eta _{\rm H}=\left\{
 \begin{array}{ll}
 \dot{M}_{\rm cr}/|\dot{M}_{\rm 2}|, & |\dot{M}_{\rm 2}|> \dot{M}_{\rm
 cr},\\
 1, & \dot{M}_{\rm cr}\geq |\dot{M}_{\rm 2}|\geq\frac{1}{8}\dot{M}_{\rm
 cr},\\
 0, & |\dot{M}_{\rm 2}|< \frac{1}{8}\dot{M}_{\rm cr}.
\end{array}\right.
\end{equation}

When a certain amount of helium is accumulated, helium is ignited
as a He-flash, and some of the helium is blown off from the
surface of the CO WD. Then, the mass growth rate of the CO WD,
$\dot{M}_{\rm WD}$, is
 \begin{equation}
 \dot{M}_{\rm WD}=\eta_{\rm He}\dot{M}_{\rm He}=\eta_{\rm He}\eta_{\rm
 H}|\dot{M}_{\rm 2}|,
  \end{equation}
where $\eta_{\rm He}$ is the mass accumulation efficiency for
helium-shell flashes, and its value is taken from \citet{KH2004}.
We assume that if an accreting CO WD increases its mass to 1.378
$M_{\odot}$ (\cite{NTY84}), it explodes as a SN Ia.



\section{Results}
\label{sect:3}
\subsection{Final State of Binary Evolution}
\label{sect:3.1}

\begin{figure}
  \begin{center}
    \includegraphics[width=6.cm, angle=270]{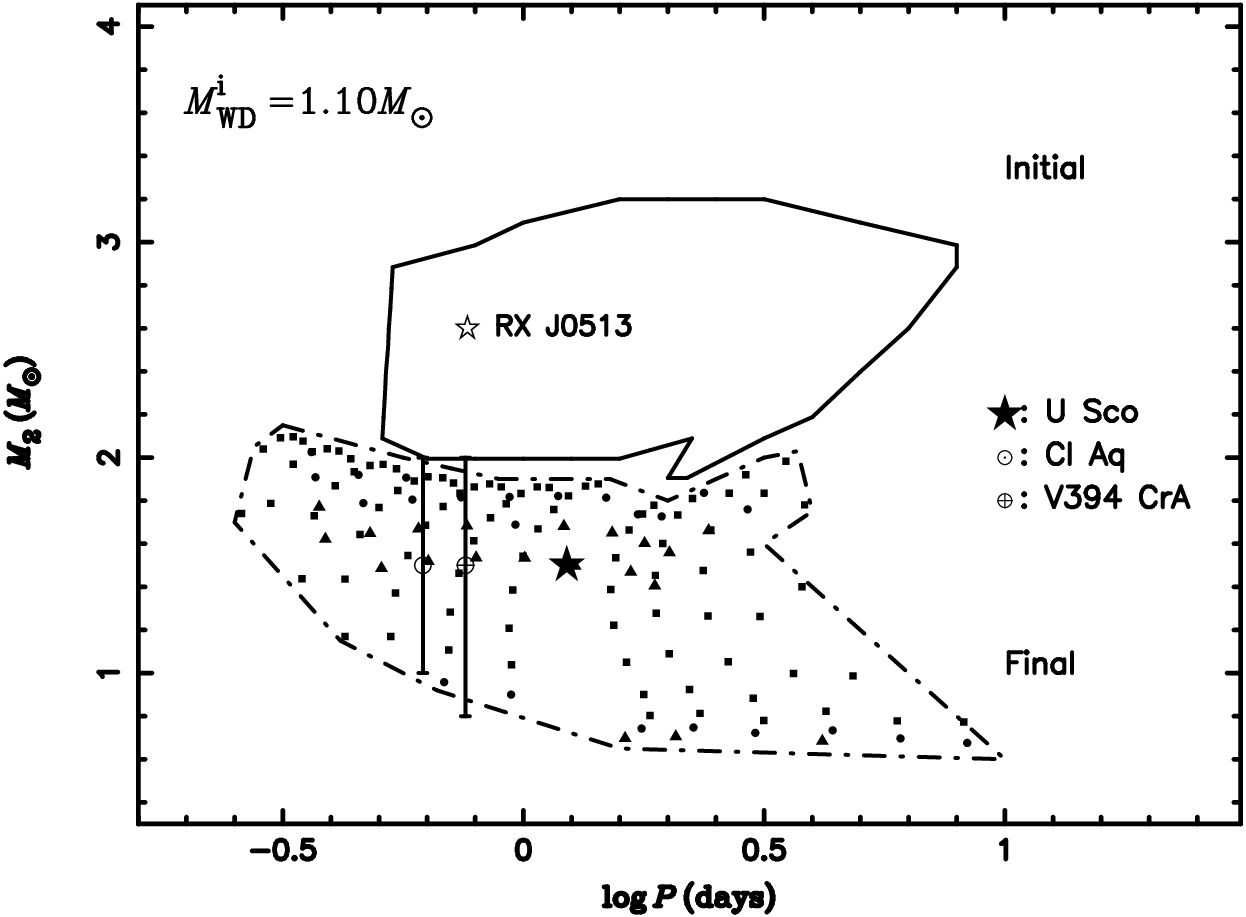}
  \end{center}
  \caption{Parameter regions producing SNe Ia in the ($\log P-M_{\rm 2}$)
  (orbital period-donor mass) plane for the WD
+ MS systems. The initial WD mass is $1.10 M_{\odot}$. The WD + MS
system insides the region encircled by solid line (labeled
``initial'') will increase its white dwarf mass up to 1.378
$M_{\odot}$, where we assume a SN Ia explosion. The final state of
the WD + MS system in the plane is encircled by dot-dashed line
(labeled ``final''). Filled squares indicate SN Ia explosions
during an optically thick wind phase ($|\dot{M}_{\rm
2}|>\dot{M}_{\rm cr}$). Filled circles denote SN Ia explosions
after the wind phase, where hydrogen-shell burning is stable
($\dot{M}_{\rm cr}\geq |\dot{M}_{\rm 2}|\geq
\frac{1}{2}\dot{M}_{\rm cr}$). Filled triangles denote SN Ia
explosions after the wind phase where hydrogen-shell burning is
mildly unstable ($\frac{1}{2}\dot{M}_{\rm cr}> |\dot{M}_{\rm
2}|\geq \frac{1}{8}\dot{M}_{\rm cr}$). A supersoft X-ray sources,
RX J0513.9-6951 (open star) is plotted, whose orbital period is
0.763 days (\cite{PAKULL93}). Three recurrent novae are indicated
by a filled star for U Sco whose period is 1.2306 days
(\cite{SR95}; \cite{HAC00a, HAC00b}), a solar symbol for CI Aq
whose orbital period is 0.6184 days (\cite{MH95}) and an earth
symbol for V394 CrA whose orbital period is 0.7577 days
(\cite{SCHAEFER90}).} \label{mms02110}
\end{figure}

   \begin{figure}[h!!!]
   \centering
   \includegraphics[width=6.cm, angle=270]{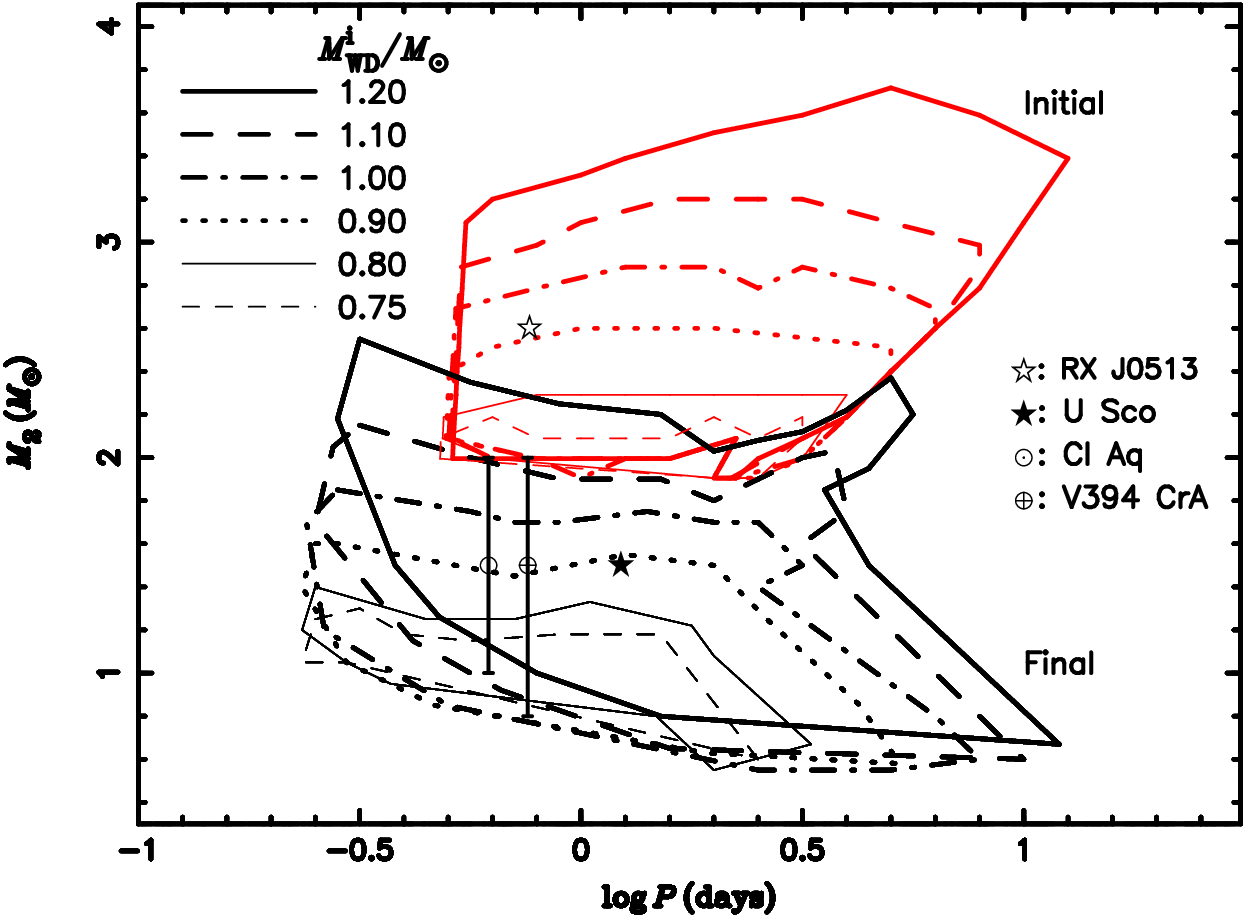}


   \caption{Similar to Fig. \ref{mms02110}, but for different initial WD mass.}
   \label{mms02}
   \end{figure}

   \begin{figure}[h!!!]
   \centering
   \includegraphics[width=6.cm, angle=270]{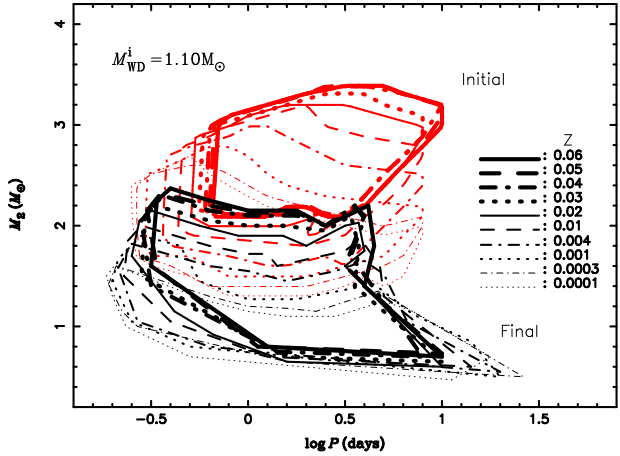}


   \caption{Similar to Fig. \ref{mms02110}, but for different metallicities.}
   \label{mmsz110}
   \end{figure}

In figure \ref{mms02110}, we show the initial contour for SNe Ia
and the final state of binary evolution in ($\log P-M_{\rm 2}$)
plane at the moment of SNe Ia explosion for the case of $M_{\rm
WD}^{\rm i}=1.1M_{\odot}$ with $Z=0.02$. The accreting WD may
reach a mass of 1.378 $M_{\odot}$ during the optically thick wind
phase (the filled squares, Case Wind) or after optically thick
wind while in stable (the filled circles, Case Calm) or unstable
(the filled triangles, Case Nova) hydrogen-burning phase. For the
first glance, the position of the final contour is much lower than
that of the initial contour, which results from the mass transfer
from secondary to white dwarf and the mass loss from the system. A
SSS, RX J0513.9-6951 (open star), whose WD mass is 1.30
$M_{\odot}$ (\cite{HK03b}), locates in the initial contour. It
should be a good candidate for the progenitor of SN Ia, as
suggested in \citet{HK01} and \citet{HK03b}. Three recurrent novae
(U Sco, V394 CrA and CI Aql) are outside the initial contour for
SNe Ia, while still inside the final region. The WD mass of U Sco
was estimated to be 1.37 $M_{\odot}$ (\cite{HAC00a, HAC00b}),
which is very close to 1.378 $M_{\odot}$. The MS mass of U Sco  is
1.5 $M_{\odot}$(\cite{HAC00a,HAC00b}), which still has enough
material to transfer onto the WD and increases the WD mass to
1.378 $M_{\odot}$. So, U Sco is a very likely candidate for the
progenitor of SN Ia (see also \cite{HAC00a, HAC00b}). The WD
masses of V394 CrA and CI Aql were estimated to be 1.37
$M_{\odot}$ and 1.2 $M_{\odot}$, respectively. Their companion
masses are still unclear. The best-fit companion masses for the
two recurrent novae are 1.50 $M_{\odot}$, while
$1.0M_{\odot}\sim2.0M_{\odot}$ is still accepted
(\cite{HK00,HK03a}; \cite{HKS03}). So, based on our binary
evolution calculation, V394 CrA is a very likely progenitor of SN
Ia and CI Aql is a possible progenitor of SN Ia. A further
observation to confirm their companion masses is necessary for
judging their fates. Another very famous recurrent nova is V 1974
Cygni (Nova Cygni 1992), whose companion mass is 0.21 $M_{\odot}$
(\cite{DEYOUNG94}; \cite{RETTER97}) and orbital period is 0.0813
days (\cite{PARESCE95}; \cite{RETTER97}). Its position is outside
the final contour in this paper. Considering its low WD mass ($1.0
M_{\odot}\sim1.1M_{\odot}$, \cite{HK05}), it is very unlikely for
V 1974 Cygni to become a SN Ia (See \citet{HKN08} for similar
discussions about these recurrent novae).

In figure \ref{mms02}, we present the initial contours for SNe Ia
and the final state in ($\log P-M_{\rm 2}$) plane at the moment of
SNe Ia explosion for different initial WD masses with $Z=0.02$.
This figure further confirms the likelihood for V394 CrA and CI Aq
to be the progenitors of SNe Ia.

In figure \ref{mmsz110}, we show the trend of the initial and
final contour with metallicity. For clarity, we only show the case
of $M_{\rm WD}^{\rm i}=1.10 M_{\odot}$, since the other cases give
similar results (see Fig. 4 in \cite{MEN09a}). We see from the
figure that there is a clear trend for the initial and final
contour to move to higher mass with metallicity. This is due to
the correlation between stellar evolution and metallicity.
Generally, stars with high metallicity evolve in a way similar to
those with low metallicity but less mass (\cite{UME99};
\cite{CHE07}; \cite{MEN08}). Thus, for binaries of CO WDs with
particular orbital periods, the companion mass increases with
metallicity (see \cite{MEN09a} in details).

\subsection{Case Wind}
\label{sect:3.2}

   \begin{figure}[h!!!]
   \centering
   \includegraphics[width=6.0cm, angle=270]{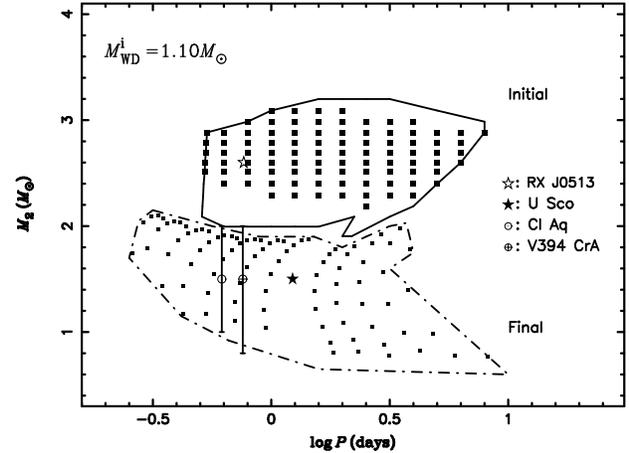}


   \caption{Similar to figure \ref{mms02110}, but only for Case Wind. The initial WD mass is 1.10 $M_{\odot}$. The big field squares show
   the initial WD + MS systems, while the small ones show their final positions at the moment of
   SNe Ia explosion. }
   \label{wind02110}
   \end{figure}

   \begin{figure}[h!!!]
   \centering
   \includegraphics[width=6.0cm, angle=270]{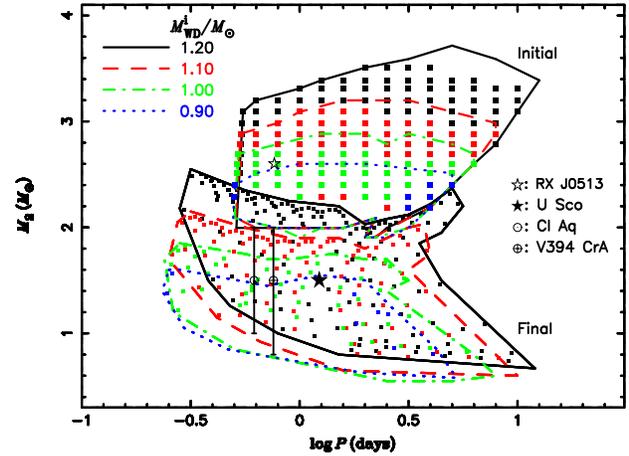}


   \caption{Similar to figure \ref{wind02110}, but for different initial WD mass. Different colors represent
   different initial WD masses. The big field squares show
   the initial WD + MS systems, while the small ones show their final positions at the moment of
   SNe Ia explosion. In the figure, we do not plot the cases with $M_{\rm i}<0.9M_{\odot}$
   for the clarity of appearance of the figure since no Case Wind occurs for the cases.}
   \label{wind02}
   \end{figure}

We present the initial and final state of binary evolution of SNe
Ia for Case Wind in figures \ref{wind02110} and \ref{wind02}.
Figure \ref{wind02110} is only for the case of $M_{\rm WD}^{\rm
i}=1.1M_{\odot}$ and figure \ref{wind02} is for different initial
WD masses. Here, we only show the case of $Z=0.02$ as a typifier,
since the other cases give similar results (see figure
\ref{mmsz110}). In these two figure, we see that the range of
initial MS masses with $Z=0.02$ for the Case Wind is between 2.2
$M_{\odot}$ and 3.5 $M_{\odot}$. However, the range strongly
depends on metallicity, that is: the progenitor systems have more
massive companions for a higher Z (see figure \ref{mmsz110} in
this paper and Fig. 4 in \cite{MEN09a}). \citet{HKN08} found that
when $M_{\rm WD}^{\rm i}\leq0.9M_{\odot}$, no Case Wind exists for
$Z=0.02$. We get a similar results for $Z=0.02$. However, this
result also rests on metallicity. For example, the mass limit for
no wind is 0.8 $M_{\odot}$ for $Z=0.06$, while it is larger than
1.0 $M_{\odot}$ for $Z=0.0001$. In addition, \citet{HKN08} noticed
that only when $M_{\rm 2}^{\rm i}\geq3.0M_{\odot}$, WDs explode at
the optically thick wind phase as SNe Ia. Their result is much
different from that in this paper. The difference is derived from
the different treatment on mass transfer between WD and MS and
from different mass-loss mechanism from systems. Firstly, they
used a analytic method for treating binary interaction. This
method can not describe certain mass-transfer phase, in particular
those occurring on a thermal time-scale and may overestimate
mass-transfer phase on the thermal time-scale (\cite{LAN00}; see
also the Figs. 1 and 4 in \cite{HAN04}). Secondly, they assume a
mass-stripping effect, that is: optically thick winds from WD
collide with secondary and strip off its surface layer. This
effect may attenuate the mass-transfer rate between WD and its
companion, but the total mass-loss rate from secondary increases
(\cite{HK03a,HK03b}; \cite{HKN08}). For the two reasons above,
more materials thus are lost from binary systems as the optically
thick wind and stripped-off material before supernova explosion,
and then the systems are more likely to explode after the
optically thick wind.

The material lost as the optically think wind forms circumstellar
material (CSM) which may result in a color excess of SNe Ia
(\cite{MEN09b}). Since the CSM is very close to SN Ia for Case
Wind, the ejecta of explosion may interact with the CSM, and the
SN Ia might then be observed as supernovae like SN 2006X which
show a variable Na I D line (\cite{PAT07}; \cite{BLONDIN09}). We
will discuss this in the future paper.

In the final contours, there is no special region for Case Wind
(see figure \ref{wind02}), that is: WDs can explode as SNe Ia at
any position in the final permitted region in the ($\log P-M_{\rm
2}$) plane, which is much different from the Case Calm and the
Case Nova (see the following subsection).

\subsection{Case Calm}
\label{sect:3.3}

   \begin{figure}[h!!!]
   \centering
   \includegraphics[width=6.0cm, angle=270]{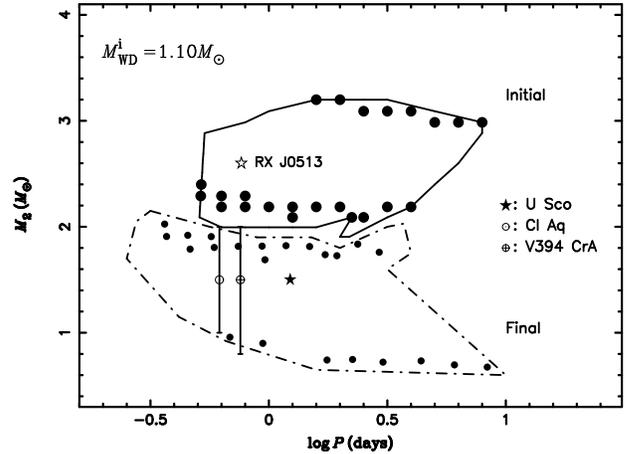}


   \caption{Similar to figure \ref{wind02110}, but only for Case Calm. The big field circles show
   the initial WD + MS systems, while the small ones show their final positions at the moment of
   SNe Ia explosion. }
   \label{sss02110}
   \end{figure}

   \begin{figure}[h!!!]
   \centering
   \includegraphics[width=6.0cm, angle=270]{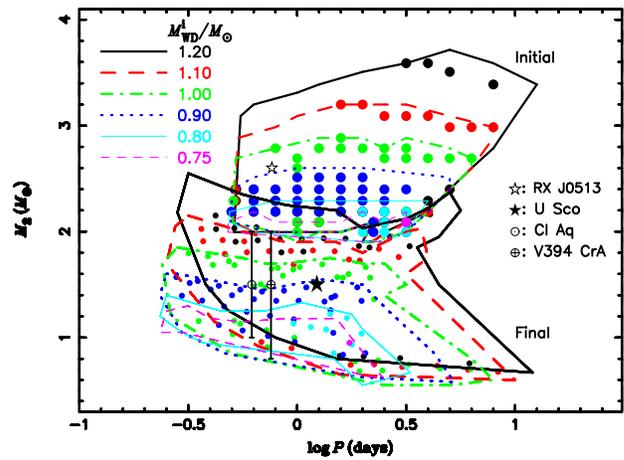}


   \caption{Similar to Fig. \ref{sss02110}, but for different initial WD masses. Different colors represent
   different initial WD masses. The big field circles show
   the initial WD + MS systems, while the small ones show their final positions at the moment of
   SNe Ia explosion.}
   \label{sss02}
   \end{figure}

If the mass-transfer rate from secondary onto WD is below the
critical accretion rate (Equation 2) while above the lowest rate
of steady hydrogen burning before supernova explosion, i.e.
$\dot{M}_{\rm cr}\geq |\dot{M}_{\rm 2}|\geq
\frac{1}{2}\dot{M}_{\rm cr}$, the WD undergoes steady H burning at
the moment of SN Ia explosion. This is the reason why we call this
``Case Calm'' (Same to that in \cite{HKN08}). During the steady
hydrogen burning phase, the WD may be observed as a SSS. The
material lost as the optically thick wind forms CSM, but it has
been dispersed too washy to be detected immediately after the SN
Ia explosion. In figures \ref{sss02110} and \ref{sss02}, we show
the initial and final state of binary evolution of SNe Ia for Case
Calm in the ($\log P-M_{\rm 2}$) plane. Figure \ref{sss02110} is
only for the case of $M_{\rm WD}^{\rm i}=1.1M_{\odot}$ and figure
\ref{sss02} is for different initial WD masses. Here, we only show
the case of $Z=0.02$ as a typifier, since the other cases give
similar results (see also figure \ref{mmsz110}). In figure
\ref{sss02110}, it is clear that there is gap for both initial
system and final state in the ($\log P-M_{\rm 2}$) plane, and the
gap divides the Case Calm into two groups. The gap in the initial
contour is the region for Case wind and the area of the gap
decrease with $M_{\rm WD}^{\rm i}$ decreasing. When $M_{\rm
WD}^{\rm i}<0.9M_{\odot}$, the gap disappear (see figure
\ref{sss02}).

The gap is also shown in the initial ($\log P-M_{\rm 2}$) plane of
\citet{HAN04} and \citet{MEN09a}, but they do not discussed it. In
this paper, we explain its origin as followings. For the high-mass
group in the initial contour of figure \ref{sss02110}, their mass
ratio is large. After the onset of Roche lobe overflow (RLOV),
mass-transfer rate is so high that the system almost undergoes
dynamically unstable. At this stage, a large amount of
hydrogen-rich materials lose as the optically thick wind and hence
the secondary mass decreases sharply. The mass-transfer rate drops
after the mass ratio has been reversed and then the optically
thick wind stops. When CO WD mass increases to 1.378 $M_{\odot}$,
the secondary mass decreases to about 0.9 $M_{\odot}$, which
corresponds to the low-mass group in the final state contour of
figure \ref{sss02110}.

For the low-mass group in the initial contour of figure
\ref{sss02110}, their mass ratio is not very large and only a
small amount of hydrogen-rich materials lose as the optically
thick wind, which means that most of transferred materials are
accumulated on the WD. Then, the secondary mass decreases
slightly, which forms the high-mass group in the final state
contour of figure \ref{sss02110}.

In figure \ref{sss02}, there is a blank region in the final state
region for Case Calm (around $M_{\rm 2}^{SN}=1.4M_{\odot}$-$\log
P=0.6$), which means that SSSs may not be observed at this region
before SNe Ia explosion. This phenomena is much different from
that of Case Wind. The blank region is mainly derived from the
low-mass group in the initial contour. For a system belonging to
the group, mass transfer is almost conservative, and then the
period is always decreasing since mass ratio is not reversed,
which leads to the region absenting long-period systems. For a
system in the high-mass, although its period decrease during
optically thick wind, the period increases after mass-ratio
reversion since no material loses from system and mass transfer
becomes conservative (see the panel (h) of Fig.1 in \cite{HAN04}).
Thus, its final period is a bit complicated.

\subsection{Case Nova}
\label{sect:3.4}

   \begin{figure}[h!!!]
   \centering
   \includegraphics[width=6.0cm, angle=270]{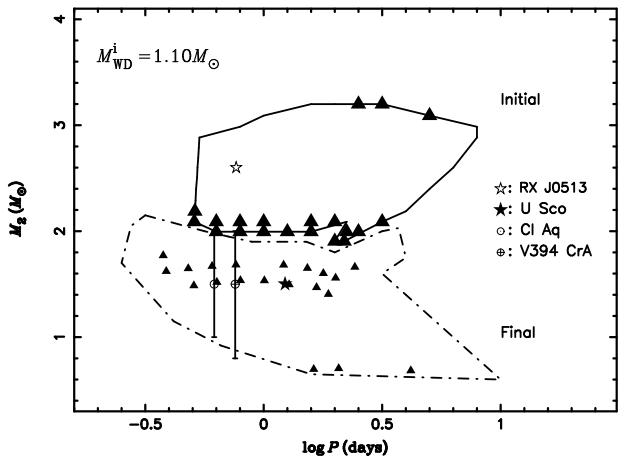}


   \caption{Similar to figure \ref{wind02110}, but only for Case Nova. The big field triangles show
   the initial WD + MS systems, while the small ones show their final positions at the moment of
   SNe Ia explosion.}
   \label{nova02110}
   \end{figure}

   \begin{figure}[h!!!]
   \centering
   \includegraphics[width=6.0cm, angle=270]{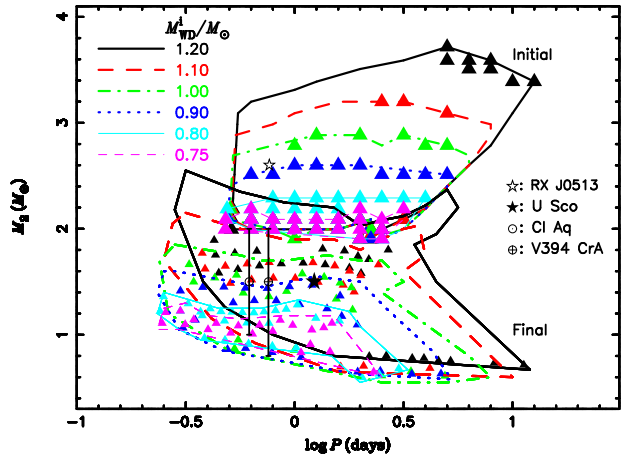}


   \caption{Similar to figure \ref{nova02110}, but for different initial WD masses. Different colors represent
   different initial WD masses. The big field circles show
   the initial WD + MS systems, while the small ones show their final positions at the moment of
   SNe Ia explosion.}
   \label{nova02}
   \end{figure}

If mass-transfer rate from secondary onto WD is below the lowest
rate of steady hydrogen burning, while still higher than a special
value, i.e. $\frac{1}{2}\dot{M}_{\rm cr}> |\dot{M}_{\rm 2}|\geq
\frac{1}{8}\dot{M}_{\rm cr}$, hydrogen shell burning is unstable
to flash, and this can recur many times in a short period as a
recurrent nova. We call this ``Case Nova'' (similar to
\cite{HKN08}). During the last recurrent nova phase, the amount of
the hydrogen-rich material lost from system is very small. On the
other hand, the material lost as the optically thick wind is
already much far from the center of SN Ia explosion. It will take
100-1000 yr for the SN Ia ejecta to catch the material lost as the
optically thick wind (\cite{HKN08}), and then no signal about CSM
can be detected immediately after supernova explosion. In figures
\ref{nova02110} and \ref{nova02}, we show the initial and final
state of binary evolution of SNe Ia for Case Nova in the ($\log
P-M_{\rm 2}$) plane. Figure \ref{nova02110} is only for the case
of $M_{\rm WD}^{\rm i}=1.1M_{\odot}$ and figure \ref{nova02} is
for different initial WD masses. Here, we only show the case of
$Z=0.02$ as a typifier, since the other cases give similar results
(see also figure \ref{mmsz110}). The initial region of Case Nova
is also divided into two group, which is similar to that of Case
Calm. The systems located in the gap are those for Case Wind or
Case Calm. The difference between Case Calm and Case Nova is that
the gap disappear when $M_{\rm WD}^{\rm i}=0.8M_{\odot}$, not
0.9$M_{\odot}$ (see figure \ref{nova02}). For a similar reason to
that of Case Calm, the high-mass group in the initial contour
forms the low-mass group in the final contour, while the low-mass
group in the initial contour forms the high-mass group in the
final contour.

Similar to figure \ref{sss02}, there is also a blank region in
figure \ref{nova02} (around $M_{\rm 2}^{SN}=1.3M_{\odot}$-$\log
P=0.5$). The origin of the blank region is similar to that of Case
Calm.

The recurrent nova U Sco, which is an excellent candidate of SN Ia
progenitor (\cite{HAC00a,HAC00b}), is located in the middle of the
final region, and its position in the ($\log P-M_{\rm 2}$) plane
is just the permitted region for recurrent nova (see figure
\ref{nova02}). Our model can explain the position of U Sco
excellently.

\subsection{Companion state after SNe Ia explosion.}
\label{sect:3.5}

   \begin{figure}[h!!!]
   \centering
   \includegraphics[width=6.0cm, angle=270]{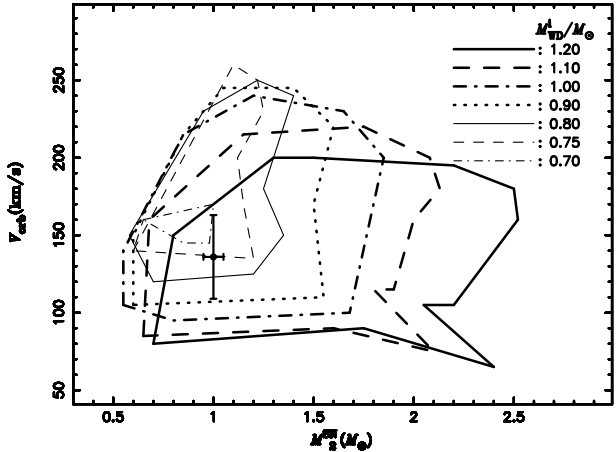}


   \caption{The final state of companion star at the moment of SNe Ia explosion for different initial WD mass with $Z=0.02$.
   Cross represents Tycho G, which is a potential candidate for the
companion of Tycho's supernova (\cite{RUI04}; \cite{BRA04}). The
length of the cross represents observational error.}
   \label{mv02}
   \end{figure}

   \begin{figure}[h!!!]
   \centering
   \includegraphics[width=6.0cm, angle=270]{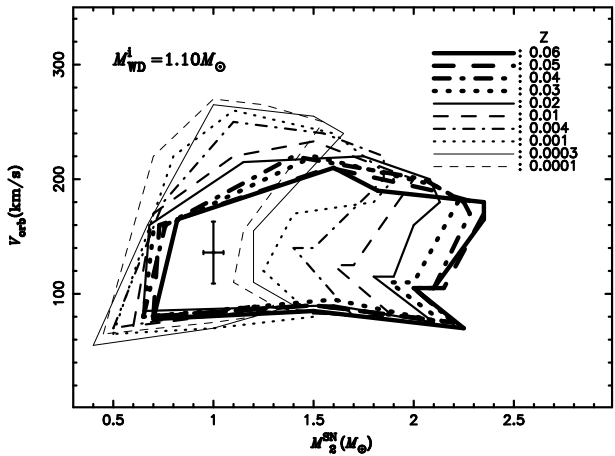}


   \caption{Similar to figure \ref{mv02}, but for different metallicities with $M_{\rm WD}^{\rm i}=1.10M_{\odot}$.}
   \label{mvz110}
   \end{figure}

A good way of discriminating between the many SN Ia progenitor
scenarios is to search for the companion of a SN Ia in its
remnant. Unless the companion is another WD (DD channel, in which
it has been destroyed by the mass-transfer process itself before
explosion), it survives and shows some special properties in its
spectra, which is originated from the contamination of supernova
ejecta (\cite{MAR00}; \cite{RUI04}; \cite{BRA04}). Tycho's
supernova, which is one of only two SNe Ia observed in our Galaxy,
provides an opportunity to address observationally the
identification of the surviving companion. \citet{RUI04} searched
the region of the remnant of Tycho's supernova and suggested that
Tycho G, a sun-like star, is the companion of Tycho's supernova.
Although, \citet{IHA07} argued that the spectrum of Tycho G do not
show any special properties, which seems to exclude the
possibility of Tycho G to be the companion of Tycho' supernova,
the analysis of the chemical abundances of the Tycho G upholds the
companion nature of the Tycho G (\cite{HERNANDEZ09}).

However, the knowledge about the companions of SNe Ia after
explosions is still unclear. Generally, the supernova ejecta in
the single degenerate model collides into the envelope of its
companion and strips some hydrogen-rich material from the surface
of the companion. After the collision, the companion gains a kick
velocity, which is much smaller than orbital velocity, and leaves
explosion center at a velocity similar to its orbital velocity
(\cite{MAR00}; \cite{MEN07}).

For a companion star in the remnant of a SN Ia, its mass and space
velocity can be detected directly. Since the stripped-off
hydrogen-rich material from companion surface and the change of
the space velocity of companion resulting from the collision of
explosion ejecta are both small (\cite{MAR00}; \cite{MAT05};
\cite{LEO07}; \cite{MEN07}), we can approximately use the state of
a companion at the moment of supernova explosion to present its
final state after supernova explosion. In figure \ref{mv02}, the
final states of companion stars are presented in $V_{\rm
orb}-M_{\rm 2}^{\rm SN}$ (orbital velocity-final companion mass)
plane for different initial WD mass with $Z=0.02$. From the
figure, we can see that the range of space velocity is from 70
km/s to 210 km/s and the mass is between 0.6 $M_{\odot}$ and 2.5
$M_{\odot}$. Tycho G just locates in this range. Our work can
explain the observation. In the future paper, we will give
detailed results about Tycho G by the approach of binary
population synthesis.

The effect of metallicity on the final state of the companions are
shown in figure \ref{mvz110}. In the figure, we only show the case
of $M_{\rm WD}^{\rm i}=1.10M_{\odot}$, since the other cases give
similar results. we can see from the figure that all the
metallicity may explain the properties of Tycho G. But for a
detailed binary population synthesis study, only the case of
$Z=0.02$ can well explain the position of Tycho G in the $V_{\rm
orb}-M_{\rm 2}^{\rm SN}$ plane. Please notice our following paper.


\section{Discussion}
\label{sect:4}
\subsection{Initial and Final Contour}
\label{sect:4.1}

In this paper, we show the initial and final parameter spaces for
SNe Ia in ($\log P-M_{\rm 2}$) plane. The final masses of the
companions are between 0.6 $M_{\odot}$ and 2.5 $M_{\odot}$ for
$Z=0.02$. However, the range of the companion mass in
\citet{HKN08} is from 1.2 $M_{\odot}$ to about 3 $M_{\odot}$. This
difference is directly derived from the different initial
contours, which is mainly determined by the different treatment of
thermal timescale mass transfer and different mass-loss mechanism.
They use a simply analytic method to estimate the mass-transfer
rate, which leads to a different initial contour for SNe Ia,
especially for low-mass WD (\cite{HAN04}). In addition, they
assumed a mass-stripping effect, and then, the initial companion
mass may be as large as 8 $M_{\odot}$, which depends on the
efficiency of the mass-stripping effect. Since their final mass of
companion is always larger than 1.2 $M_{\odot}$, it is difficult
for their model to explain the properties of Tycho G, which
remains consistent with the surviving companion of Tycho's
supernova (1 $M_{\odot}$, \cite{RUI04}; \cite{HERNANDEZ09}). Since
our results are based on detailed binary evolution calculation
with latest input physics, our results are more realistic and the
final contours in this paper are more likely to approach the real
ones than those in \citet{HKN08} (see also in \cite{HAN04}).
Because of the difference range of final companion as mentioned
above, Tycho G can be a likely progenitor of Tycho's supernova in
our model, but can not be in the model of \citet{HKN08}.

\subsection{Supernova like SN 2002ic and V Sge}
\label{sect:4.2}

   \begin{figure}[h!!!]
   \centering
   \includegraphics[width=6.0cm, angle=270]{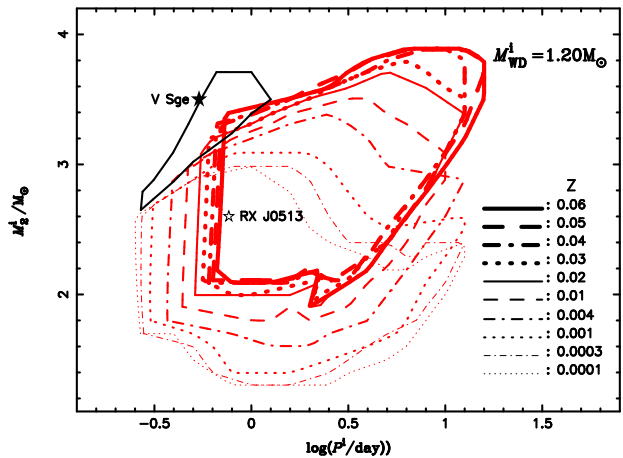}


   \caption{Initial region for SNe Ia with different metallicities. Here, only the cases of $M_{\rm WD}^{\rm i}=1.20M_{\odot}$
   are shown. The black solid line shows the initial region of the progenitor for supernovae like SN 2002ic.
   Two supersoft X-ray sources, RX J0513.9-6951 (open star) and V Sge (filled star), are plotted, whose masses are 2.7 $M_{\odot}$
   and 3.5 $M_{\odot}$ (\cite{HK03b, HK03c}), and whose orbital period are 0.76 days and 0.51 days, respectively (\cite{PAKULL93};
   \cite{HERBIG65}; \cite{PATTERSON98}).}
   \label{120}
   \end{figure}

Until SN 2002ic was discovered (\cite{HAM03}), it was long
believed that there are no hydrogen lines in the spectra of SNe
Ia. The strong hydrogen lines in the spectra of SN 2002ic were
explained by the interaction between the SN ejecta and the
circumstellar material (CSM) (\cite{HAM03}).  Recently, two
co-twins of SN 2002ic (SN 2005gj and SN 2006gy) were also found
(\cite{ALD06}; \cite{OFE07}). To explain these rare objects, many
models were suggested and here we just list some of them:
\citet{HAC99b}; \citet{HAM03}; \citet{LR03}; \citet{CHU04};
\citet{HAN06}; \citet{HKN08}; \citet{LGL09}. Among all the models
listed above, the delayed dynamical instability model suggested by
\citet{HAN06} is more interesting and some predictions from the
model seem to be consistent with observations, especially in a
sense of the birth rate and delay time of these rare objects
(\cite{ALD06}; \cite{PRI07b}). In the scenario of \citet{HAN06},
SN 2002ic might be from the WD + MS channel, where the CO WD
accretes material from its relatively massive companion ($\sim
3.0M_{\odot}$), and increases its mass to $\sim 1.30 M_{\odot}$
before experiencing a delayed dynamical instability. Adopting
their model, \citet{MEN09a} assumed that the CO WD can increase
its mass to 1.378 $M_{\odot}$ and explode as a SN 2002ic-like case
if the mass of the CO WD exceeds 1.30 $M_{\odot}$ before the
delayed dynamical instability, and then predicted that supernovae
like SN 2002ic may not be found in extremely low-metallicity
environments. Under the same assumption to \citet{MEN09a}, we show
the permitted region for the delayed dynamical instability model
with various metallicities in figure \ref{120}. For the simplicity
of the assumption, the real region for the delayed dynamical
instability model may be larger than that shown in figure
\ref{120} (see Fig. 5 in \cite{HAN06} and Fig. 2. in
\cite{MEN09a}). It is clear that the permitted region for
supernova like SN 2002ic locates above the contours for normal
supernova. There is a overlap between supernova like SN 2002ic and
normal SN Ia, since a progenitor of supernova like SN 2002ic with
low metallicity may have a lower mass than that of normal
supernova. Whatever, more evidence is necessary to confirm the
scenario in \citet{HAN06}, especially to find a progenitor system
at present.

V Sge is a well observed quasi-periodic transient SSS in our
Galaxy. It switches optical state between a high ($V\sim11$ mag)
and low ($V\sim12$ mag) state, and during the low state, very soft
and very weak X-ray can be detected (\cite{GvT98}). The mass-loss
rate for V Sge can be as large as $\sim10^{\rm -5}M_{\odot}{\rm
yr^{\rm -1}}$ indicated by radio observation.
(\cite{LOCKLEY97,LOCKLEY99}). \citet{HK03c} use optically thick
wind to explain the observational property of V Sge excellently,
and in their model, the mass-loss rate may reaches as high as
$\sim10^{\rm -5}M_{\odot}{\rm yr^{\rm -1}}$, which is consistent
with radio observations. \citet{HK03c} also estimated the WD mass
as $M_{\rm WD}\sim1.25M_{\odot}$ and its companion mass as $M_{\rm
2}\sim3.5M_{\odot}$, and then they suggested that V Sge will
explode as a SN Ia after about $10^{\rm -5}{\rm yr}$. However,
they must assume a mass-stripping effect, otherwise the position
of V Sge in ($\log P-M_{\rm 2}$) plane will beyond the permitted
region for stable mass transfer (\cite{HKN08}). We get a similar
result to that in \citet{HKN08} that V Sge locates in a forbidden
zone for stable mass transfer if there is no special assumption
(see in figure \ref{120}). However, it locates at the boundary of
the contour for supernovae like SN 2002ic as indicated in
\citet{HAN06}. Considering the smaller region here than a real one
and a moderately larger WD mass of V Sge than that in this paper,
V Sge is a potential progenitor system that will undergo the
delayed dynamical instability as suggested in \citet{HAN06}. V Sge
thus would not explode as a normal SN Ia, but one like SN 2002ic,
if the model in \citet{HAN06} is appropriate. Then, we might
regard ``V Sge-type star" as the progenitor of supernovae like SN
2002ic if it belongs to WD + MS system.

The lifetime of V Sge-type star is typically $\sim10^{\rm 5}{\rm
yr}$, which is mainly due to the time-averaged wind mass-loss rate
of $\sim10^{\rm -5}M_{\odot}{\rm yr^{\rm -1}}$ (\cite{HAN06}).
Since no more than 1 in 100 SNe Ia belong to the subgroup of
2002ic-like supernova (\cite{HAN06}; \cite{ALD06}; \cite{PRI07b}),
there should be several V Sge-type star belonging to WD + MS
system in our Galaxy if we take the Galactic birth rate of SNe Ia
as 3-4$\times10^{\rm -3}{\rm yr^{\rm -1}}$ (\cite{VAN91};
\cite{CT97}). \citet{STEINER98} listed four V Sge-type stars in
the Galaxy and discussed their similar spectroscopic and
photometric properties. At present, their companion masses have
not been identified, while their orbital periods is in the range
of 0.2-0.5 days, which falls into the orbital periods predicted by
our model (see the final regions in figure \ref{mmsz110}). Please
notice that not all SSSs belong to WD + MS systems and the nature
of the compact star still remains an open question
(\cite{STEINER98}). White dwarfs, neutron stars, black holes, or
even carbon main-sequence stars are all possible candidates
(\cite{STEINER98}). Another alternative for the SSSs is WD + He
star system, which may also contribute to a few part of SNe Ia
(\cite{WANGBO09}). Thus, considering the uncertainty of the nature
of the four V Sge-type star, our estimation on the number of V
Sge-type stars is consistent with observations.

However, although the life, the mass-lose rate and the number of V
Sge-type star seem to match with the prediction from the delayed
dynamical instability model, the case of $M_{\rm WD }^{\rm
i}=1.2M_{\odot}$ is the only one in our model grids to account for
the position of V Sge in the ($\log P-M_{\rm 2}$) plane, while
there exist arguments on whether the mass of a CO WD may be as
large as $1.2M_{\odot}$ and whether the CO WD of $1.2M_{\odot}$
may explode as a SN Ia (\cite{UME99}; \cite{MEN08}). So, obtaining
a conclusion on whether there is a relation between V Sge-type
star and the delayed dynamical instability model is premature, and
V Sge is only a potential candidate for the progenitor of
supernovae like SN 2002ic.

\section{Summary and Conclusions}
\label{sect:5}

Adopting the prescription in \citet{HAC99a} for the mass accretion
of CO WDs, we have carried out detailed binary evolution
calculations for the progenitors of SNe Ia with different
metallicity in the single-degenerate channel (the WD + MS channel)
and obtained the initial and final parameters in the ($\log
P-M_{\rm 2}$) plane. Our model may explain the positions of some
famous recurrent novae in the ($\log P-M_{\rm 2}$) plane, as well
as a SSS, RX J0513.9-6951. Our model can also explain the space
velocity and mass of Tycho G, which is now suggested to be the
potential companion star of Tycho's supernova (\cite{RUI04};
\cite{HERNANDEZ09}).  Based on the delayed dynamically instable
model in \citet{HAN06}, we might regard V Sge as a potential
progenitor of supernova like SN 2002ic.



\begin{thebibliography}{}

\bibitem[Aldering et al.(2006)]{ALD06}
Aldering G., Antilogus P., Bailey S. et al., 2006, ApJ, 650, 510
\bibitem[Alexander \& Ferguson(1994)]{AF94}
Alexander D. R., Ferguson J. W., 1994, ApJ, 437, 879
\bibitem[Arnett(1982)]{ARN82}
Arnett W.D., 1982, ApJ, 253, 785
\bibitem[Blondin et al.(2009)]{BLONDIN09}
Blondin S., Prieto J.L., Patat F. et al., 2009, ApJ, 693, 207
\bibitem[Branch et al.(1995)]{BRANCH95}
Branch D., Livio M., Yungelson L.R. et al., 1995, PASP, 107, 1019
\bibitem[Branch(2004)]{BRA04}
Branch D., 2004, Nature, 431, 1044
\bibitem[Branch(2006)]{BRA06}
Branch D., 2006, Nature, 443, 283
\bibitem[Cappellaro \& Turatto(1997)]{CT97}
Cappellaro E., Turatto M., 1997, in Ruiz-Lapuente P., Cannal R.,
Isern J., eds, Thermonuclear Supernovae. Kluwer, Dordrecht, p. 77
\bibitem[Chugai \& Yungelson(2004)]{CHU04}
Chugai N.N., Yungelson L.R., 2004, Astronomy Letters, 30, 65
\bibitem[Chen \& Tout(2007)]{CHE07}
Chen X., Tout C.A., 2007, ChJAA, 7, 2, 245
\bibitem[Deyoung \& Schmidt(1994)]{DEYOUNG94}
Deyoung J. A. \& Schmidt R.E., 1994, ApJ, 431, L47
\bibitem[Di Stefano \& Kong(2003)]{DIK03}
Di Stefano R., Kong A.K.H., 2003, ApJ, 592, 884
\bibitem[Eggleton(1971)]{EGG71}
Eggleton P.P., 1971, MNRAS, 151, 351
\bibitem[Eggleton(1972)]{EGG72}
Eggleton P.P., 1972, MNRAS, 156, 361
\bibitem[Eggleton(1973)]{EGG73}
Eggleton P.P., 1973, MNRAS, 163, 279
\bibitem[Greiner \& van Teeseling(1998)]{GvT98}
Greiner J. \& van Teeseling A., 1998, A\&A, 118, 217
\bibitem[Hachisu et al.(1996)]{HAC96}
Hachisu I., Kato M., Nomoto K., ApJ, 1996, 470, L97
\bibitem[Hachisu et al.(1999a)]{HAC99a}
Hachisu I., Kato M., Nomoto K., Umeda H., 1999a, ApJ, 519, 314
\bibitem[Hachisu et al.(1999b)]{HAC99b}
Hachisu I., Kato M., Nomoto K., 1999b, ApJ, 522, 487
\bibitem[Hachisu et al.(2000a)]{HAC00a}
Hachisu I., Kato M., Kato T., Matsumoto K., 2000a, ApJ, 528, L97
\bibitem[Hachisu et al.(2000b)]{HAC00b}
Hachisu I., Kato M., Kato T., Matsumoto K., Nomoto K., 2000b, ApJ,
534, L189
\bibitem[Hachisu \& Kato(2000)]{HK00}
Hachisu I., Kato M., 2000, ApJ, 540, 447
\bibitem[Hachisu \& Kato(2001)]{HK01}
Hachisu I., Kato M., 2001, ApJ, 558, 323
\bibitem[Hachisu \& Kato(2003a)]{HK03a}
Hachisu I., Kato M., 2003a, ApJ, 588, 1003
\bibitem[Hachisu \& Kato(2003b)]{HK03b}
Hachisu I., Kato M., 2003b, ApJ, 590, 445
\bibitem[Hachisu \& Kato(2003c)]{HK03c}
Hachisu I., Kato M., 2003c, ApJ, 598, 527
\bibitem[Hachisu, Kato \& Schaefer(2003)]{HKS03}
Hachisu I., Kato M., Schaefer B.E., 2003, ApJ, 584, 1008
\bibitem[Hachisu \& Kato(2005)]{HK05}
Hachisu I., Kato M., 2005, ApJ, 631, 1094
\bibitem[Hachisu \& Kato(2006a)]{HK06a}
Hachisu I., Kato M., 2006a, ApJ, 642, L52
\bibitem[Hachisu \& Kato(2006b)]{HK06b}
Hachisu I., Kato M., 2006b, ApJ, 651, L141
\bibitem[Hachisu et al.(2007)]{HKL07}
Hachisu I., Kato M., Luna G.J.M., 2007, ApJ, 659, L153
\bibitem[Hachisu et al.(2008)]{HKN08}
Hachisu I., Kato M., Nomoto K., 2008, ApJ, 679, 1390 (arXiv:
0710.0319)
\bibitem[Hamuy et al.(2003)]{HAM03}
Hamuy M. et al., 2003, Nature, 424, 651
\bibitem[Han, Podsiadlowski \& Eggleton(1994)]{HAN94}
Han Z., Podsiadlowski P., Eggleton P.P., 1994, MNRAS, 270, 121
\bibitem[Han(1998)]{HAN98}
Han Z., 1998, MNRAS, 296, 1019
\bibitem[Han et al.(2000)]{HAN00}
Han Z., Tout C.A., Eggleton P.P., 2000, MNRAS, 319, 215
\bibitem[Han \& Podsiadlowski(2004)]{HAN04}
Han Z., Podsiadlowski Ph., 2004, MNRAS, 350, 1301
\bibitem[Han \& Podsiadlowski(2006)]{HAN06}
Han Z., Podsiadlowski Ph., 2006, MNRAS, 368, 1095
\bibitem[Hern\'{a}ndez et al.(2009)]{HERNANDEZ09}
Hern\'{a}ndez J.I.G., Ruiz-lapuente P., Filippenko A.V., Foley
R.J., Gal-Yam A., Simon J.D., 2009, ApJ, 691, 1
\bibitem[Herbig et al.(1965)]{HERBIG65}
Herbig G.H., Preston G.W., Smak J., Paczy$\acute{\rm n}$ski B.,
1965, ApJ, 141, 617
\bibitem[Hillebrandt \& Niemeyer(2000)]{HN00}
Hillebrandt W., Niemeyer J.C., 2000, ARA\&A, 38, 191
\bibitem[Howell et al.(2006)]{HOW06}
Howell D.A. et al., 2006, Nature, 443, 308
\bibitem[Iben \& Tutukov(1984)]{IT84}
Iben I., Tutukov A.V., 1984, ApJS, 54, 335
\bibitem[Iglesias \& Rogers(1996)]{IR96}
Iglesias C. A., Rogers F. J., 1996, ApJ, 464, 943
\bibitem[Ihara et al.(2007)]{IHA07}
Ihara Y., Ozaki J., Doi M. et al., 2007, PASJ, 59, 811, arXiv:
0706.3259
\bibitem[Kato \& Hachisu(2004)]{KH2004}
Kato M., Hachisu I., 2004, ApJ, 613, L129
\bibitem[Kobayashi et al.(1998)]{KOB98}
Kobayashi C., Tsujimoto T., Nomoto K. et al., 1998, ApJ, 503, L155
\bibitem[Langer et al.(2000)]{LAN00}
Langer N., Deutschmann A., Wellstein S. et al., 2000, A\&A, 362,
1046
\bibitem[Leibundgut(2000)]{LEI00}
Leibundgut B., 2000, A\&ARv, 10, 179
\bibitem[Leonard(2007)]{LEO07}
Leonard D.C., 2007, ApJ, 670, 1275
\bibitem[Li \& van den Heuvel(1997)]{LI97}
Li X.D., van den Heuvel E.P.J., 1997, A\&A, 322, L9
\bibitem[Livio \& Riess(2003)]{LR03}
Livio M., Riess A., 2003, ApJ, 594, L93
\bibitem[L\"{u} et al.(2009)]{LGL09}
L\"{u} G., Zhu C. Wang Z., Wang N., 2009, MNRAS, 396, 1086,
arXiv:0903.2636
\bibitem[Lockley et al.(1997)]{LOCKLEY97}
Lockley J.J., Eyres S.P.S., Wood J.H., 1997, MNRAS, 287, L14
\bibitem[Lockley et al.(1999)]{LOCKLEY99}
Lockley J.J., Wood J.H., Eyres S.P.S., Naylor T., Shugarov S.,
1999, MNRAS, 310, 963
\bibitem[Marietta et al.(2000)]{MAR00}
Marietta E., Burrows A., Fryxell B., 2000, ApJS, 128, 615
\bibitem[Mattila et al.(2005)]{MAT05}
Mattila S., Lundqvist P., Sollerman J. et al., 2005, A\&A, 443,
649
\bibitem[Meng et al.(2006)]{MEN06}
Meng X., Chen X., Tout C.A., Han Z., 2006, ChJAA, 6, 4, 461
\bibitem[Meng et al.(2007)]{MEN07}
Meng X., Chen X., Han Z., 2007, PASJ, 59, 835
\bibitem[Meng, Chen \& Han(2008)]{MEN08}
Meng X., Chen X., Han Z., 2008, A\&A, 487, 625, arXiv: 0710.2397.
\bibitem[Meng, Chen \& Han(2009)]{MEN09a}
Meng X., Chen X., Han Z., 2009, MNRAS, 395, 2103, arXiv:0802.2471
\bibitem[Meng et al.(2009)]{MEN09b}
Meng X., Chen X., Han Z., Yang W., 2009, RA\&A, accepted,
arXiv:0907.2753
\bibitem[Mennickent \& Honeycutt(1995)]{MH95}
Mennickent R.E. \& Honeycut R.K., 1995, Inf. Bull.Variable Stars,
4232
\bibitem[Nomoto, Thielemann \& Yokoi(1984)]{NTY84}
Nomoto K., Thielemann F-K., Yokoi K., 1984, ApJ, 286, 644
\bibitem[Nomoto \& Kondo (1991)]{NK91}
Nomoto K., Kondo Y., 1991, ApJ, 367, L19
\bibitem[Nomoto et al.(1999)]{NOM99}
Nomoto K., Umeda H., Hachisu I. Kato M., Kobayashi C., Tsujimoto
T., 1999, in Truran J., Niemeyer T., eds, Type Ia Suppernova
:Theory and Cosmology.Cambridge Univ. Press, New York, p.63
\bibitem[Nomoto et al.(2003)]{NOM03}
Nomoto K., Uenishi T., Kobayashi C. Umeda H., Ohkubo T., Hachisu
I., Kato M., 2003, in Hillebrandt W., Leibundgut B., eds, From
Twilight to Highlight: The Physics of supernova, ESO/Springer
serious ``ESO Astrophysics Symposia'' Berlin: Springer, p.115
\bibitem[Ofek et al.(2007)]{OFE07}
Ofek E.O., Cameron P.B., Kaslwal M.M. et al., 2007, ApJ, 659, L13,
arXiv: 0612408
\bibitem[Pakull et al.(1993)]{PAKULL93}
Pakull M.W., Moch C., Bianchi L., Thomas H.C., Guibert J.,
Beaulieu J.P., Grison P., \& Schaeidt S., 1993, A\&A, 278, L39
\bibitem[Paresce et al.(1995)]{PARESCE95}
Paresce F., Livio M., Hack W., Korista K., 1995, A\&A, 299, 823
\bibitem[Parthasarathy et al.(2007)]{PAR07}
Parthasarathy M., Branch D., Jeffery D.J., Baron E., 2007, NewAR,
51, 524, arXiv: 0703415
\bibitem[Patat et al.(2007)]{PAT07}
Patat E. et al., Science, 317, 924
\bibitem[Patterson et al.(1998)]{PATTERSON98}
Patterson J., Kemp J., Shambrook A. et al., 1998, PASP, 110, 380
\bibitem[Perlmutter et al.(1999)]{PER99}
Perlmutter S. et al., 1999, ApJ, 517, 565
\bibitem[Pols et al.(1995)]{POL95}
Pols O.R., Tout C.A., Eggleton P.P. et al., 1995, MNRAS, 274, 964
\bibitem[Pols et al.(1997)]{POL97}
Pols O.R., Tout C.A., Schr\"{o}der K.P. et al., 1997, MNRAS, 289,
869
\bibitem[Pols et al.(1998)]{POL98}
Pols O.R., Schr\"{o}der K.P., Hurly J.R. et al., 1998, MNRAS, 298,
525
\bibitem[Prieto et al.(2008)]{PRI07a}
Prieto J.L. et al., 2008, ApJ, 673, 999, arXiv: 0707.0690
\bibitem[Prieto et al.(2007)]{PRI07b}
Prieto J.L. et al., 2007, arXiv: 0706.4088
\bibitem[Quimby, H\"{o}flich \& Wheeler(2007)]{QUI07}
Quimby R., P. H\"{o}flich, J.C. Wheeler, 2007, ApJ, 666, 1083
\bibitem[Retter, Leibowitz \& Ofek(1997)]{RETTER97}
Retter A., Leibowitz E.M. \& Ofek E.O., 1997, MNRAS, 286, 745
\bibitem[Riess et al.(1998)]{RIE98}
Riess A. et al., 1998, AJ, 116, 1009
\bibitem[Ruiz-Lapuente et al.(2004)]{RUI04}
Ruiz-Lapuente P. et al., 2004, Nature, 431, 1069
\bibitem[Schr\"{o}der et al.(1997)]{SCH97}
Schr\"{o}der K.P., Pols O.R., Eggleton P.P., 1997, MNRAS, 285, 696
\bibitem[Schaefer \& Ringwald(1995)]{SR95}
Schaefer B.E. \& Ringwald F.A., 1995, ApJ, 447, L45
\bibitem[Schaefer(1990)]{SCHAEFER90}
Schaefer B.E., 1990, ApJ, 355, L39
\bibitem[Shanks et al.(2002)]{SHA02}
Shanks T., Allen P.D., Hoyle F. et al., 2002, ASPC, 283, 274
\bibitem[Steiner \& Diaz(1998)]{STEINER98}
Steiner J.E. \& Diaz M.P., 1998, PASP, 110, 276
\bibitem[Umeda et al.(1999)]{UME99}
Umeda H., Nomoto K., Yamaoka H. et al., 1999, ApJ, 513, 861
\bibitem[van den Bergh \& Tammann(1991)]{VAN91}
van den Bergh S., Tammann G.A., 1991, ARA\&A, 29, 363
\bibitem[Voss \& Nelemans(2008)]{VOSS08}
Voss R., \& Nelemans G., 2008, Nature, 451, 802
\bibitem[Wang et al.(2009)]{WANGBO09}
Wang, B.; Meng, X.; Chen, X.; Han, Z., 2009, MNRAS, in press,
arXiv0901.3496
\bibitem[Whelan \& Iben(1973)]{WI73}
Whelan J., Iben I., 1973, ApJ, 186, 1007
\bibitem[Whelan \& Iben(1987)]{WI87}
Whelan J., Iben I., 1987, in Philipp A.G.D., Hayes D.S., Liebert
J.W., eds, IAU Colloq.95, Second Conference on Faint Blue Stars.
Davis Press, Schenectady, p. 445
\bibitem[Yungelson et al.(1995)]{YUN95}
Yungelson L., Livio M., Tutukou A. Kenyon S.J., 1995, ApJ, 447,
656

\end{thebibliography}
\end{document}